\documentclass[12pt,dvips]{article}
\usepackage[dvips]{graphics}
\def\numt#1#2{#1 \times 10^{#2}}
\def\fglab#1{\label{fig:#1}}
\def\PL#1#2#3{Phys. Lett. {\bf #1}, #2 (#3)}

\begin{document}
\baselineskip13pt

\begin{center}
{\Large \bf 
The quantum effects on neutrino mass and lepton 
 flavor mixing (MNS) 
 matrices\footnote{Talk at Post Summer Institute 2000, 
 Yamanashi, Japan, 21 Aug - 24 Aug 2000.}
}\\
\vglue 3mm
Naoyuki Haba\footnote{E-mail: haba@eken.phys.nagoya-u.ac.jp}\\
\vglue 3mm
{\small \it $^2$Faculty of Engineering, Mie University,}
{\small \it Tsu Mie 514-8507, Japan}\\
\end{center}
\noindent{\large \bf 1. Introduction}\\

\noindent
Previously, people thought the effects of  
 quantum corrections on neutrino mass matrix 
 may be negligible, since the Yukawa couplings of
 neutrinos are much much smaller than other quarks 
 and leptons. 
However, is it true??
Besides, are ``maximal'' mixings in the lepton flavor 
 mixing matrix stable against quantum corrections?
My talk is concentrating on these topics 
 in the framework of the minimal supersymmetric
 standard model with the effective dimension-five 
 operator $\kappa_{ij}$ which
 gives the Majorana masses of neutrinos. \\

\noindent{\large \bf 2. RGE of neutrino sector}\\

\noindent
The renormalization group equations (RGEs) of neutrino 
 mass and MNS matrices are given by 
 simple formulas thanks to 
 the non-renormalization theorem of supersymmetry. 
The RGE of the operator $\kappa_{ij}$ is given by 
\begin{equation}
{d \over d t} \kappa_{ij} = 
 ( \gamma_i + \gamma_j + 2 \gamma_H ) \kappa_{ij}
\end{equation}
where, $\gamma_{i,H}$ is the anomalous dimensions 
 of $i$-th generation 
 lepton and Higgs doublets, respectively. 
This equation induces the following 
 two important consequences\cite{HOMS}. \\
\noindent
\hspace*{1cm}$\star$ None of phases of $\kappa_{ij}$ 
 depend on the energy scale. \\
\noindent
\hspace*{1cm}$\star$ The energy scale dependence of 
 the MNS matrix is governed 
 only by \\
\hspace*{1.4cm}$n_g-1$ real parameters. \\
\noindent
Here $n_g$ is the generation number. 
Above two consequences make the RGE analyses of 
 neutrino mass and MNS matrices be 
 simple. 
Let us show the three-generation case. 
The neutrino mass matrix 
 at the high energy $m_{\nu}(M_R)$ is 
 related to that at the low energy 
 $m_{\nu}(m_Z)$ as \cite{E}\cite{HO}
\begin{equation}
 m_{\nu}(M_R)\; = \; c \;
\left(
\begin{array}{ccc}
 1- \epsilon_e &  0 &  0 \\
 0 & 1- \epsilon_{\mu}  &  0 \\
 0 & 0  &  1
\end{array}
\right) \;
m_{\nu}(m_Z) \;
\left(
\begin{array}{ccc}
 1- \epsilon_e &  0 &  0 \\
 0 & 1- \epsilon_{\mu}  &  0 \\
 0 & 0  &  1
\end{array}
\right),
\end{equation}
where $\epsilon_{e, \mu}$ is the quantities
 determined by the values of $\tan \beta$ and 
 $M_R$ as shown in Fig.1. 
Since the difference of the magnitudes between 
 $Y_e$ and $Y_{\mu}$ comparing to 
 $Y_{\tau}$ is negligible, 
 the quantum effects 
 can be estimated by 
 only one parameter 
 $\epsilon (\simeq \epsilon_{e, \mu})$\cite{E}\cite{HO}. 
\begin{figure}[htb]
\begin{center}
 {\scalebox{.5}{\includegraphics{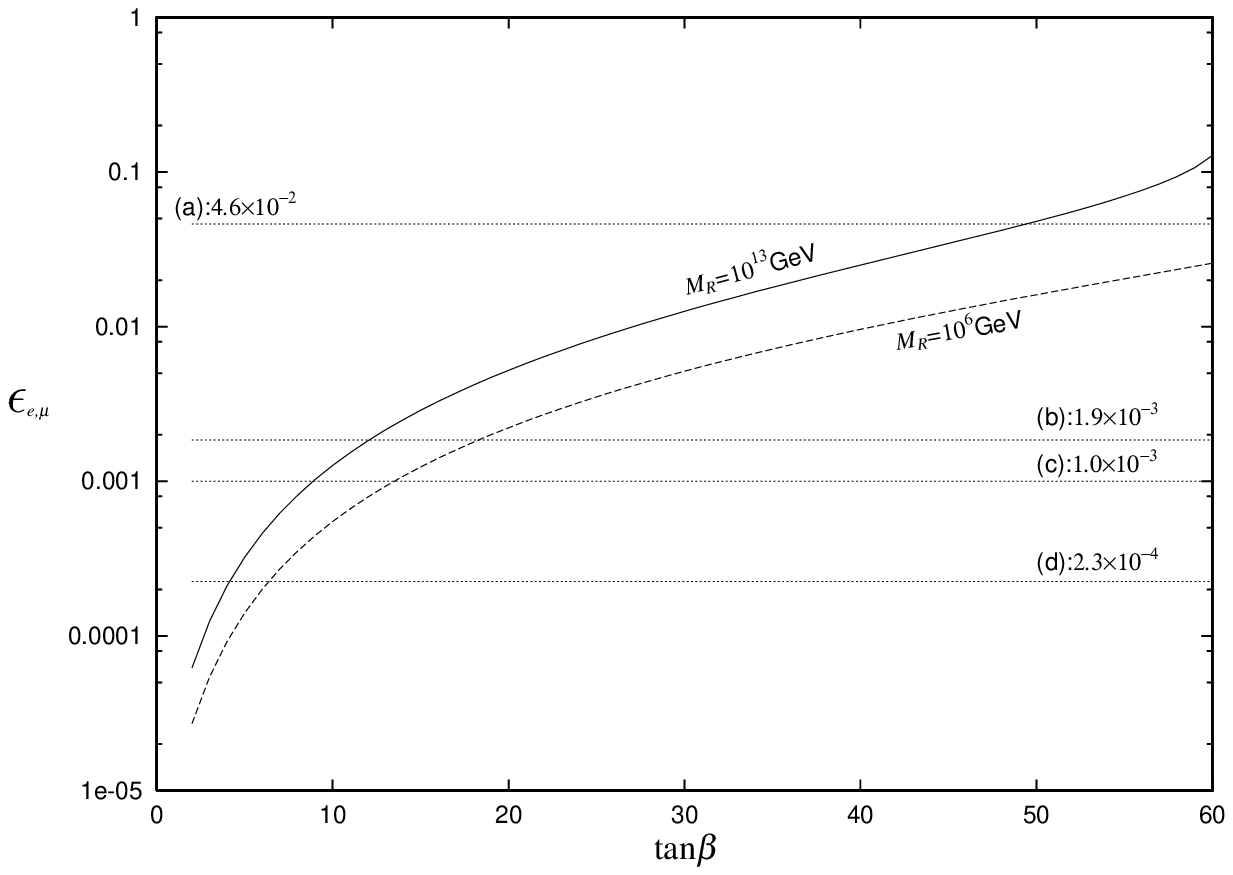}}} 
\end{center}
 \vspace{-1.5em}
 \caption{$\tan \beta$ dependence of $\epsilon_{e,\mu}$[3].
The solid-line (dashed-line) shows
$M_R^{}=10^{13}$GeV ($10^6$GeV).
Each dotted-line shows 
 (a):$\numt{4.6}{-2}$, (b):$\numt{1.9}{-3}$,
 (c):$\numt{1.0}{-3}$ and (d):$\numt{2.3}{-4}$.
}
 \fglab{epsilon_emu}
\end{figure}

Now let us show quantum effects on neutrino 
 mass and MNS matrices in $2 \times 2$ case, 
 $3 \times 3$ case, and democratic-type of mass 
 matrices.  \\

\noindent{\large \bf 3. $2 \times 2$ case}\\

\noindent
Two-generation example is the good study for 
 understanding ``more complicated'' three-generation case. 
The results of the three-generation case are 
 completely understood in the analogy of two-generation case. 
There are two cases for the maximal mixings, 
 one is $(a):$ hierarchical, and the other is 
 $(b):$ degenerate. 
{}For the first order, 
 these cases are classified as  
 $m_{\nu}^{(a)}=diag.(0,1)$, 
 $m_{\nu}^{(b1)}=diag.(-1,1)$, and 
 $m_{\nu}^{(b2)}=diag.(1,1)$ 
 in the diagonal base of neutrino mass 
 matrix. 
Here $(b1)$ and $(b2)$ are cases with 
 opposite signs. 
We can easily show that the mixing angle of 
 cases $(a)$ and 
 $(b1)$ are stable, and $(b2)$ is unstable 
 against quantum corrections\cite{HO}. 
The case of {\it degenerate} and 
{\it the same sign} has a risk of 
 instability of mixing angles\footnote{It depends
 on the values of Majorana mass and
 $\tan \beta$\cite{HO}.}. 
In other words, a ``maximal'' mixing is possibly 
 obtained by the RGE effects in the case of $(b2)$.
We had shown a ``maximal'' mixing can be realized 
 by the quantum corrections in degenerate 
 neutrinos with (the same sign) 
 masses of order 0.1 eV\cite{HOS}.

\par
We can also show that 
 the cases of $(b1)$ and $(b2)$ are connected 
 with each other by the 
 physical Majorana $CP$ phase $\phi$. 
Therefore, this $CP$ phase should be the order
 parameter connecting the ``stable'' region with the 
 ``unstable'' region as shown in Fig.2\cite{HOMS2}. 
\begin{figure}[htb]
\begin{center}
 {\scalebox{.3}{\includegraphics{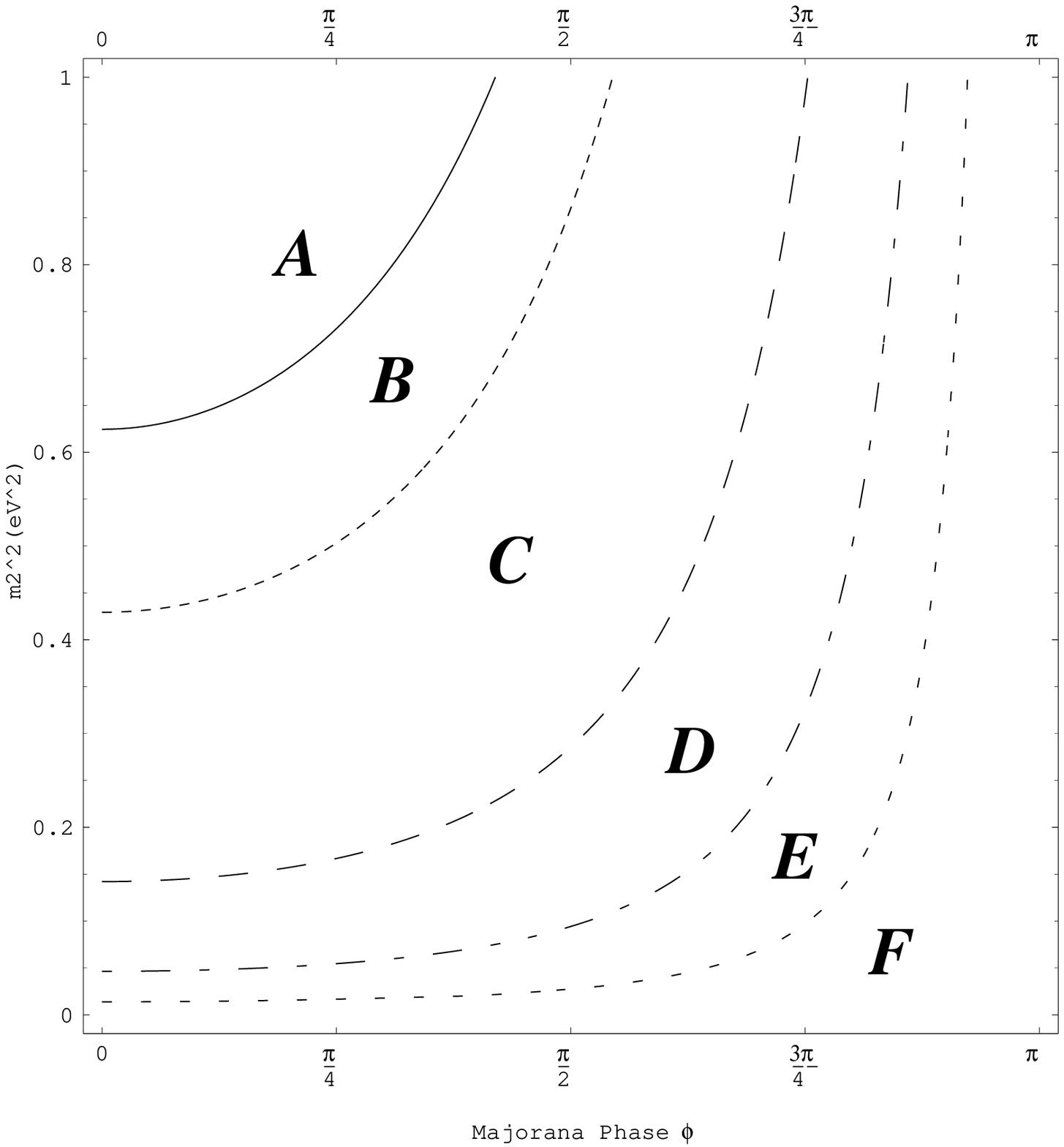}}} 
 \vspace{-1.5em}
 \caption{contour plot of $\sin^2 2 \theta_{23}$ for 
 $\phi$ and $m_2^2$[5].
{\small \hspace{1cm}$A: \sin^2 2 \bar{\theta_{23} < 0.05} $, 
$B: 0.05 \leq \sin^2 2 \bar{\theta_{23} < 0.1}$, 
$C: 0.1 \leq \sin^2 2 \bar{\theta_{23} < 0.5}$, 
$D: 0.5 \leq \sin^2 2 \bar{\theta_{23} < 0.9}$, 
$E: 0.9 \leq \sin^2 2 \bar{\theta_{23} < 0.99}$, 
$F: 0.99 \leq \sin^2 2 \bar{\theta_{23}}$.}} 
\end{center}
 \fglab{cp02}
\end{figure}
%

\noindent{\large \bf 4. $3 \times 3$ case}\\

\noindent
The stabilities of mixing angles in the 
 $3 \times 3$ case can be understood
 in the analogy of $2 \times 2$ case. 
The results are shown in Ref.\cite{HO}, 
 that is, cases of 
 {\it hierarchical} or 
 {\it the opposite signs} between the generations 
 guarantee stable mixing angles 
 against quantum corrections. 
As for the $CP$ phases, there are two
 physical Majorana phases, which connect 
 the ``stable'' region with the ``unstable'' 
 region\cite{HOM} as in 
 the case of $2 \times 2$ case.  \\

\noindent{\large \bf 5. Stability of the 
 democratic-type of mass matrix}\\

\noindent
Democratic type 
 of mass matrix\footnote{In order to realize 
 the democratic-type of mass matrix, 
 there should be, for example, $S_3{}_L \times S_3{}_R$, 
 $O(3)_L \times O(3)_R$ flavor symmetries behind.} 
 can naturally 
 explain why masses of the third generation fields 
 are much heavier than those of other generation fields. 
In the democratic-type of mass matrix, 
 the origin of the ``maximal''mixing exists in 
 the unitary matrix diagonalyzing the charged
 lepton sector, where the neutrino masses should 
 be degenerate with the same signs and negligibly 
 small flavor mixings. 
As shown in $2 \times 2$ case of $(b2)$, 
 this case has a risk 
 that the mixing angles become 
 unstable due to quantum corrections. 
Actually, stabilities of the mixing angles 
 depend on the solar solutions (degrees of degeneracy). 
When we take the right-handed Majorana mass scale 
 as $10^{13}$ GeV, the vacuum solution is 
 completely destroyed by the quantum corrections, 
 and the large angle and the small angle 
 MSW solutions require $\tan \beta < 10$ in order not to 
 be destroyed by the quantum corrections.  \\

\noindent{\large \bf 6. Summary}\\

\noindent
The quantum effects of lepton flavor 
 mixing angles can be easily estimated by 
 using the technique in Ref.\cite{HOMS}. 
The cases of {\it hierarchical} or 
 {\it the opposite signs} between the generations 
 guarantee stable mixing angles 
 against quantum corrections. 
We also show that physical Majorana 
 phases connect the ``stable'' region with the 
 ``unstable'' region.


\begin{thebibliography}{1}

\bibitem{HOMS}
N. Haba, Y. Matsui, N. Okamura and M. Sugiura,
Eur. Phys. J. C10, 677 (1999).

\bibitem{E}
J. Ellis and S. Lola,
\PL{B458}{310}{1999}.


\bibitem{HO}
N. Haba and N. Okamura, 
Eur. Phys. J. C14, 347 (2000).

\bibitem{HOS}
N. Haba, N. Okamura and M. Sugiura,
Prog. Theor. Phys. 103, 367 (2000).


\bibitem{HOMS2}
N. Haba, Y. Matsui, N. Okamura and M. Sugiura,
Prog. Theor. Phys. 103, 807 (2000).


\bibitem{HOM}
N. Haba, Y. Matsui and N. Okamura, 
hep-ph/0005075, Eur. Phys. J. in press. 


\end{thebibliography}
\end{document}